\documentclass[%
 reprint,
 superscriptaddress,
 nofootinbib,
 amsmath,amssymb,
 aps,
 pra,
]{revtex4-1}

\usepackage{graphicx}
\usepackage{color}

\begin{document}

\title{All-loop singularities of scattering amplitudes in massless planar theories}

\author{Igor Prlina}
\affiliation{Department of Physics, Brown University, Providence RI 02912}

\author{Marcus Spradlin}
\affiliation{Department of Physics, Brown University, Providence RI 02912}
\affiliation{School of Natural Sciences, Institute for Advanced Study, Princeton NJ 08540}

\author{Stefan Stanojevic}
\affiliation{Department of Physics, Brown University, Providence RI 02912}

\begin{abstract}
In massless quantum field theories the Landau equations
are invariant under graph operations familiar from the theory of
electrical circuits.
Using a theorem on the $Y$-$\Delta$ reducibility of planar
circuits we prove that the set of first-type Landau singularities of
an $n$-particle scattering amplitude in any massless planar theory,
in any spacetime dimension $D$,
at any finite loop order in perturbation theory,
is a subset of those of a certain $n$-particle
$\lfloor{(n{-}2)^2/4}\rfloor$-loop
``ziggurat'' graph.
We determine this singularity locus explicitly for $D=4$ and $n=6$ and find
that it corresponds precisely to the vanishing of the symbol letters
familiar from the hexagon bootstrap in SYM theory.
Further implications for SYM theory are discussed.
\end{abstract}

\maketitle

\section{Introduction}

For over half a century much has been learned from the study of singularities
of scattering amplitudes in quantum field theory, an important class
of which are encoded in the Landau equations~\cite{Landau:1959fi}.
This paper combines two simple statements to arrive at a general
result about such singularities.
The first is based on the analogy between Feynman diagrams
and electrical circuits, which also has been long appreciated
and exploited; see for example~\cite{Mathews:1959zz,Wu:1961zz,Wu}
and chapter~18 of~\cite{BJ}.
Here we use the fact that in \emph{massless} field theories,
the sets of solutions to the Landau equations are invariant
under the elementary graph operations familiar from
circuit theory, including in particular the $Y$-$\Delta$
transformation which replaces a triangle subgraph with a tri-valent
vertex, or vice versa.
The second is a theorem of Gitler~\cite{GitlerThesis},
who proved that any \emph{planar} graph (of the type
relevant to the analysis of Landau equations, specified below)
can be $Y$-$\Delta$ reduced
to a class we call \emph{ziggurats}
(see Fig.~\ref{fig:ziggurat}).

We conclude that the $n$-particle $\lfloor{(n{-}2)^2/4}\rfloor$-loop
ziggurat graph encodes all possible first-type Landau singularities
of any $n$-particle amplitude at any finite loop order
in any massless planar theory.
Although this result applies much more generally,
our original motivation arose from
related
work~\cite{Dennen:2015bet,Dennen:2016mdk,Prlina:2017azl,Prlina:2017tvx}
on planar
$\mathcal{N} = 4$ supersymmetric Yang-Mills (SYM) theory,
for which our result has several interesting implications
which we discuss in Sec.~\ref{sec:sym}.

\section{Landau Graphs and Singularities}

We begin by reviewing the Landau equations,
which encode the constraint of locality on the singularity
structure of scattering amplitudes in perturbation theory
via Landau graphs.
We aim to connect the standard vocabulary
used in relativistic
field theory to that of network theory in order to streamline the
rest of our discussion.

In \emph{planar} quantum field theories, which will be the exclusive
focus of this paper, we can restrict our attention to plane Landau
graphs.
An \emph{$L$-loop $m$-point plane Landau graph} is
a plane graph
with $L{+}1$ faces
and $m$ distinguished vertices called \emph{terminals}
that must lie on a common face called the \emph{unbounded face}.
Henceforth we use the word ``vertex'' only for
those that are not terminals, and the word ``face''
only for the $L$ faces that are not the unbounded face.

Each edge $j$ is assigned an arbitrary orientation
and a four-component (or, more generally, a $D$-component)
(energy-)momentum vector $q_j$, the analog
of electric current.
Reversing the orientation of an edge changes the sign of the
associated $q_j$.
At each vertex
the vector sum of incoming
momenta must equal the vector sum of outgoing momenta (current
conservation).
This constraint is not applied at terminals, which are the locations
where a circuit can
be probed by connecting external sources or sinks of current.
In field theory these correspond to the momenta carried by incoming or
outgoing particles.
If we label the terminals by $a=1,\ldots,m$ (in cyclic order
around the unbounded face) and let $P_a$ denote the $D$-momentum flowing
into the graph at terminal $a$, then energy-momentum conservation
requires that $\sum_a P_a = 0$ and implies that precisely $L$ of
the $q_j$'s are linearly independent.

Scattering amplitudes are (in general multivalued) functions of the
$P_a$'s which can be expressed as a sum over all Landau graphs,
followed by a $DL$-dimensional integral
over all components of the linearly independent $q_j$'s.
Amplitudes in different quantum field theories differ
in how
the various graphs are weighed (by $P_a$- and $q_j$-dependent
factors) in that linear combination.  These
differences are indicated graphically by decorating each
Landau graph (usually in many possible ways)
with various embellishments, in which case they
are called \emph{Feynman diagrams}.
We return to this important point later, but for now we keep our
discussion as general as possible.

Our interest lies in understanding
the loci in $P_a$-space on which amplitudes may have
singularities, which are highly constrained by general physical
principles.
A Landau graph is said to have
\emph{Landau singularities of the first type}
at values of $P_a$ for which the \emph{Landau equations}~\cite{Landau:1959fi}
\begin{align}
\label{eq:onshell}
\alpha_j q_j^2 &= 0 \text{ for each edge } j, \text{ and}\\
\sum_{\text{edges } j \in \mathcal{F}} \alpha_j q_j &= 0 \text{ for each face } \mathcal{F}
\label{eq:kirk}
\end{align}
admit nontrivial solutions for
the \emph{Feynman parameters} $\alpha_j$
(that means, omitting the trivial solution where
all $\alpha_j = 0$).
In the first line we have indicated our exclusive focus
on \emph{massless} field theories by omitting a term proportional
to $m_j^2$ which would normally be present.

The Landau equations generally admit several branches
of solutions.
The \emph{leading} Landau singularities of a graph $\mathcal{G}$ are those
associated to branches
having $q_j^2 = 0$ for all $j$ (regardless of whether any
of the $\alpha_j$'s are zero).  This differs slightly from
the more conventional usage of the term ``leading'', which requires
all of the $\alpha_j$'s to be nonzero.  However, we feel that our
usage is more natural in massless theories, where it is typical to
have branches of solutions
on which $q_j^2$ and $\alpha_j$ are \emph{both} zero for certain edges $j$.
Landau singularities associated to
branches on which one or more of the $q_j^2$ are not zero (in
which case the corresponding $\alpha_j$'s must necessarily vanish)
can be interpreted as
leading singularities of a \emph{relaxed} Landau graph obtained
from $\mathcal{G}$ by contracting the edges associated to the vanishing
$\alpha_j$'s.

A graph is called \emph{$c$-connected} if it remains
connected after removal of any $c{-}1$ vertices.
It is easy to see that the set of Landau singularities
for a 1-connected graph (sometimes called a ``kissing graph''
in field theory)
is the union of Landau singularities associated
to each 2-connected component since the Landau equations
completely decouple.  Therefore, without loss of generality
we can confine our attention to 2-connected Landau graphs.

\section{Elementary Circuit Operations}
\label{sec:moves}

We refer to Eq.~(\ref{eq:kirk}) as the \emph{Kirchhoff
conditions} in recognition of their circuit analog where the $\alpha_j$'s
play the role of resistances.
The analog of the \emph{on-shell conditions}~(\ref{eq:onshell})
on the other hand
is rather mysterious,
but a very remarkable feature of massless theories is that:
\vspace{0.05cm}
\begin{center}
\boxed{
\begin{minipage}{0.95\linewidth}
The graph moves familiar from elementary
electrical circuit theory preserve the solution sets of Eqs.~(\ref{eq:onshell})
and~(\ref{eq:kirk}),
and hence, the sets of first-type Landau singularities
in any massless field theory.
\end{minipage}
}
\vspace{0.05cm}
\end{center}

\begin{figure}
\includegraphics[width=0.8\linewidth]{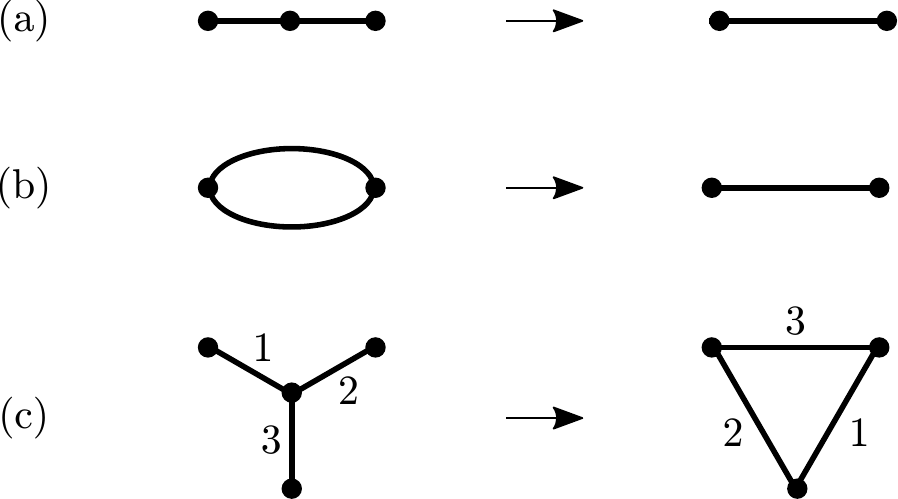}
\caption{Elementary circuit moves that preserve
solution sets of the massless Landau equations:
(a) series reduction, (b) parallel reduction, and
(c) $Y$-$\Delta$ reduction.}
\label{fig:moves}
\end{figure}

Let us now demonstrate this feature, beginning with the three
elementary circuit moves shown in Fig.~\ref{fig:moves}.

Series reduction (Fig.~\ref{fig:moves}(a))
allows one to remove any vertex of degree two.
Since $q_2 = q_1$ by momentum conservation, the structure
of the Landau equations is trivially preserved
if the two edges with Feynman parameters $\alpha_1$, $\alpha_2$
are replaced by a single edge carrying momentum
$q' = q_1 = q_2$ and Feynman parameter $\alpha' = \alpha_1 + \alpha_2$.

Parallel reduction (Fig.~\ref{fig:moves}(b))
allows one to collapse any bubble subgraph.
It is easy to verify (see for example Appendix~A.1
of~\cite{Dennen:2016mdk}) that the structure
of the Landau equations is preserved if the two edges of the bubble
are replaced by a single edge carrying momentum $q' = q_1 + q_2$
and Feynman parameter $\alpha' = \alpha_1 \alpha_2/(\alpha_1 + \alpha_2)$.

The $Y$-$\Delta$ reduction (Fig.~\ref{fig:moves}(c)) replaces a vertex of degree three (a ``$Y$'') with a triangle subgraph (a ``$\Delta$''), or vice versa.
Generically
the Feynman parameters $\alpha_i$ of the $\Delta$ are related
to those of the $Y$, which we call $\beta_i$, by
\begin{align}
\beta_1 = \frac{\alpha_2 \alpha_3}{\alpha_1 + \alpha_2 + \alpha_3}\,, \quad
\text{and cyclic.}
\end{align}
On branches where one or more of the parameters vanish, this relation
must be suitably modified.  For example, if a branch of solutions
for a graph containing a $Y$ has $\beta_1 = \beta_2 = 0$ but
$\beta_3$ nonzero, then the
corresponding branch for the reduced graph has $\alpha_3 = 0$ but
$\alpha_1, \alpha_2$ nonzero.

\begin{figure}
\includegraphics[width=0.75\linewidth]{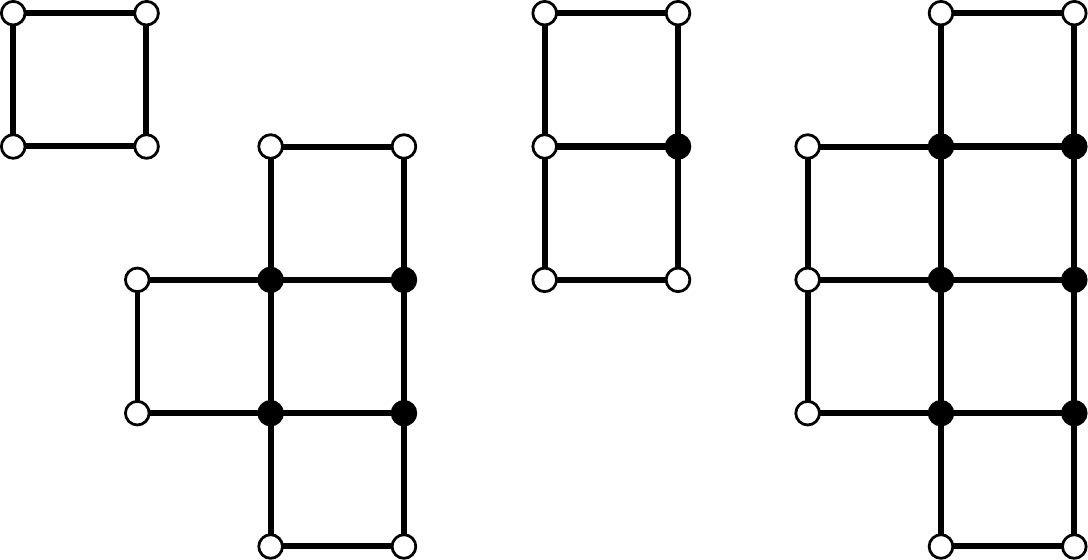}
\caption{The four-, six-, five- and seven-terminal ziggurat graphs.
The open circles are terminals and the filled circles are vertices.
The pattern continues in the obvious way, but note an
essential difference between ziggurat graphs with an even or odd number
of terminals in that only the latter have a terminal of degree three.
}
\label{fig:ziggurat}
\end{figure}

\begin{figure*}
\includegraphics[width=0.95\linewidth]{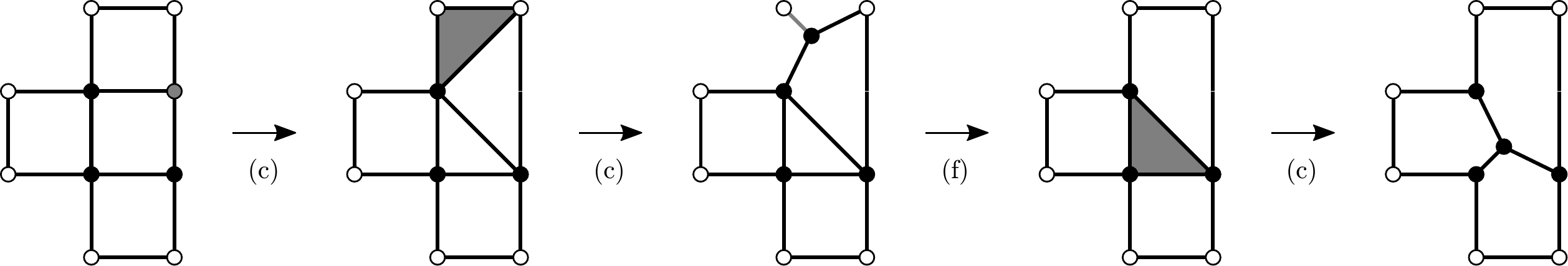}
\caption{The six-terminal ziggurat graph can be reduced to a three
loop graph by a sequence of three $Y$-$\Delta$ reductions
and one FP assignment.
In each case the vertex, edge, or face to be transformed is
highlighted in gray.}
\label{fig:reduce}
\end{figure*}

The invariance of the Kirchhoff conditions~(\ref{eq:kirk}) under
$Y$-$\Delta$ reduction follows straightforwardly from these Feynman parameter assignments.
The invariance of the on-shell conditions~(\ref{eq:onshell})
is nontrivial, and follows
from the analysis in Appendix~A.2 of~\cite{Dennen:2016mdk}
by checking that the on-shell conditions before and after the reduction
are equivalent for each branch of solutions to the Landau
equations.  Actually~\cite{Dennen:2016mdk} mentions only
seven of the eight different types of branches.
The eighth branch has
$\alpha_1 = \alpha_2 = \alpha_3 = 0$, corresponding
to $\beta_1 = \beta_2 = \beta_3 = 0$,
but in this relatively trivial case
both the $Y$ and the $\Delta$ can effectively be collapsed
to a single vertex.

The proof of the crucial theorem of~\cite{GitlerThesis} that we employ
in the next section
relies on three additional relatively simple moves that either have
no analog in field theory or trivially preserve
the essential content of the Landau equations.
These are (d) the deletion of a ``tadpole'' (edges that connect a vertex
or terminal to itself), (e) the deletion of a ``hanging propagator''
(a vertex of degree one and the edge connected to it), and (f)
the contraction of an edge connected to a terminal of degree
one
(called ``FP assignment''~\cite{Gitler:2011}).
The last of these is strictly speaking not completely trivial at the level
of the Landau equations; it just removes an otherwise uninteresting
bubble singularity.

\section{Reduction of Planar Graphs}
\label{sec:reduction}

The reduction of general graphs under the operations reviewed in the previous section is a well-studied problem in the mathematical literature.
When it is declared that a certain subset of vertices are
to be considered terminals (which may not be
removed by series or $Y$-$\Delta$ reduction)
the corresponding problem
is called \emph{terminal $Y$-$\Delta$ reducibility}.
Aspects of terminal $Y$-$\Delta$ reducibility
have been studied
in~\cite{Akers:1960,Feo:1993,Verdiere:1996,Archdaecon:2000,Gitler:2011,Demasi:2014},
including
an application to Feynman diagrams
in~\cite{Suzuki:2011hfa}.
For our purpose the key result
comes from the Ph.D.\ thesis
of I.~Gitler~\cite{GitlerThesis}, who proved that any
planar
2-connected graph with $m$ terminals
lying on the same face can be reduced to a graph
of the kind shown in Fig.~\ref{fig:ziggurat}, which we call
\emph{ziggurat} graphs,
or to a minor thereof.
We denote the $m$-terminal ziggurat graph by $\mathcal{T}_m$, and
note that
a \emph{minor} of a graph $\mathcal{G}$
is any graph that can be obtained from
$\mathcal{G}$
by any sequence of edge contractions and/or edge deletions.

At the level of Landau equations an edge contraction
corresponds, as discussed above, to a relaxation (setting
the associated $\alpha_j$ to zero), while an edge
deletion corresponds to setting the associated $q_j$ to zero.
It is clear that the Landau singularities associated
to any minor of a graph $\mathcal{G}$ are a subset of those
associated to $\mathcal{G}$.
Consequently we don't need to worry about explicitly enumerating
all minors of $\mathcal{T}_m$; their Landau singularities are already
contained in the set of singularities of $\mathcal{T}_m$ itself.

It is conventional
to discuss scattering amplitudes for a fixed number
$n$ of external particles, each of which
carries some momentum $p_i$ that in massless theories satisfies
$p_i^2 = 0$.
The total momentum flowing
into each terminal is not arbitrary, but must
be a sum of one or more null vectors.
The momenta carried by these individual particles
are denoted graphically by attaching a total of $n$
\emph{external edges} to the
terminals, with at least one per terminal.
In this way it is
clear that
any Landau graph with $m \le n$ terminals is potentially relevant
to finding the Landau singularities of an $n$-particle amplitude.
However, it is also clear that if $m < n$ then
$\mathcal{T}_m$ is a minor of
$\mathcal{T}_n$, so again the Landau singularities of the
former are a subset of those of the latter.
Therefore, to find the Landau singularities of an $n$-particle
amplitude it suffices to find those of the $n$-terminal
ziggurat graph $\mathcal{T}_n$ with precisely one external edge
attached to each terminal.  We call this the \emph{$n$-particle
ziggurat graph} and finally summarize:
\vspace{0.05cm}
\begin{center}
\boxed{
\begin{minipage}{0.95\linewidth}
The  first-type  Landau singularities of an $n$-particle scattering amplitude
in any massless planar field theory are a subset of those of the
$n$-particle ziggurat graph.
\end{minipage}
}
\vspace{0.05cm}
\end{center}

While the Landau singularities of the ziggurat graph exhaust
the set of
singularities that may appear in any massless planar theory,
we cannot rule out the
possibility that in certain special theories
the actual set of singularities may be smaller because of nontrivial
cancellation between the contributions of different Landau
graphs to a given
amplitude.  We return to this important point in
Sec.~\ref{sec:sym}.

Let us also
emphasize that $Y$-$\Delta$ reduction certainly changes
the number of faces of a graph, so the above
statement does not hold at fixed loop order $L$; rather it is an all-order
relation about the full set of Landau singularities of $n$-particle
amplitudes at any finite order in perturbation theory.
Since the $n$-particle ziggurat graph has
$L = \lfloor{(n{-}2)^2/4}\rfloor$
faces, we can however predict that
a single computation at only
$\lfloor{(n{-}2)^2/4}\rfloor$-loop order suffices to expose
all possible Landau singularities
of any $n$-particle amplitude.

In fact this bound is unnecessarily high.
Gitler's theorem does not imply that ziggurat graphs
cannot be further
reduced to graphs of lower loop order, and it is easy to see
that in general this is possible.  For example, as shown
in Fig.~\ref{fig:reduce}, the six-terminal
graph can be reduced by a sequence of $Y$-$\Delta$ reductions
and one FP assignment to a particularly beautiful
three-loop wheel graph whose 6-particle avatar
we display in Fig.~\ref{fig:wheel}.
Ziggurat graphs with more than six terminals can also be further reduced,
but we have not been able to prove a lower bound on the loop
order that can be obtained for general $n$.

\section{Landau Analysis of the Wheel}
\label{sec:landau}

\begin{figure}
\includegraphics[width=0.49\linewidth]{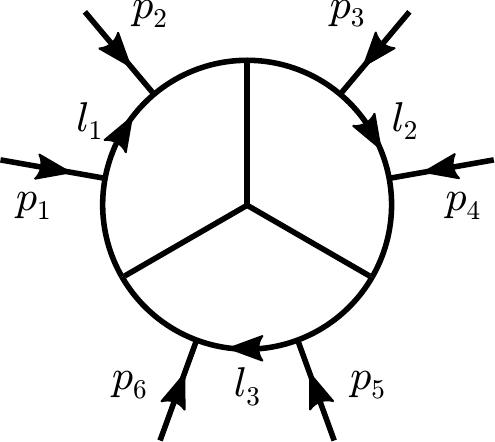}
\caption{The three-loop six-particle wheel graph.
The leading first-type Landau singularities of this graph exhaust
all possible
first-type
Landau singularities of six-particle amplitudes
in any massless planar field theory, to any finite loop order.}
\label{fig:wheel}
\end{figure}

In this section we analyze the Landau equations for
the graph shown in Fig.~\ref{fig:wheel}.
The six external edges carry momenta $p_1, \ldots, p_6$ into
the graph, subject to $\sum_i p_i = 0$ and $p_i^2 = 0$ for each $i$.
Using momentum conservation at each vertex, the momentum $q_j$
carried by each of the twelve edges
can be expressed in terms of the six $p_i$ and
three other linearly independent momenta, which we can take to be
$l_r$, for $r=1,2,3$, assigned as shown in the figure.
Initially we consider the leading Landau singularities,
for which we impose the twelve on-shell conditions
\begin{equation}
\begin{aligned}
(l_1 - p_1)^2 = l_1^2 = (l_1 + p_2)^2 = 0\,,
\\
(l_2 - p_3)^2 = l_2^2 = (l_2 + p_4)^2 = 0\,,
\\
(l_3 - p_5)^2 = l_3^2 = (l_3 + p_6)^2 = 0\,,
\\
(l_1 + p_2 - l_2 + p_3)^2 = 0\,, \\
(l_2 + p_4 - l_3 + p_5)^2 = 0 \,,\\
(l_3 + p_6 - l_1 + p_1)^2 = 0\,.
\label{eq:wheellandau}
\end{aligned}
\end{equation}
So far we have not needed to commit to any particular spacetime
dimension.
We now fix $D = 4$, which simplifies the analysis
because for generic $p_i$ there are precisely 16 discrete solutions
for the $l_r$'s, which we denote by
$l_r^{*}(p_i)$.  To enumerate and explicitly exhibit these solutions it
is technically helpful to parameterize  the momenta in terms of
momentum twistor variables~\cite{Hodges:2009hk}, in which case
the solutions can be associated with
on-shell diagrams
as described in~\cite{ArkaniHamed:2012nw}.
Although so far the analysis is still applicable to
general massless planar theories,
we note that
in the special context of SYM theory,
two cut solutions have MHV support, twelve NMHV, and
two NNMHV.

With these solutions in hand, we next turn our attention to the
Kirchhoff conditions
\begin{equation}
\begin{aligned}
0 = &\alpha_1 (l_1 - p_1) + \alpha_2 l_1 + \alpha_3 (l_1 + p_2) +\\
&\alpha_{10} (l_3 + p_6 - l_1 + p_1) + \alpha_{11} (l_1 + p_2 - l_2 + p_3)\,,  \\
0 = &\alpha_4 (l_2 - p_3) + \alpha_5 l_2 + \alpha_6 (l_2 + p_4) + \\
&\alpha_{11} (l_1 + p_2 - l_2 + p_3) + \alpha_{12} (l_2 + p_4 - l_3 + p_5)\,,  \\
0 = &\alpha_7 (l_3 - p_5) + \alpha_8 l_3 + \alpha_9 (l_3 + p_6) +\\
&\alpha_{12} (l_2 + p_4 - l_3 + p_5) + \alpha_{10} (l_3 + p_6 - l_1 + p_1)\,.
\end{aligned}
\end{equation}
Nontrivial solutions to this $12 \times 12$ linear system
exist only if the associated
\emph{Kirchhoff determinant} $K(p_i, l_r)$ vanishes.
By evaluating this determinant on each of the solutions
$l_r = l_r^*(p_i)$ the condition for the existence of a non-trivial
solution to the Landau equations can be expressed entirely in terms of the
external momenta $p_i$.
Using variables $u, v, w, y_u, y_v, y_w$ that
are very familiar in the literature on six-particle
amplitudes (their definition in terms of the
$p_i$'s can be found for example in~\cite{Caron-Huot:2016owq}),
we find that $K(p_i, l_r^*(p_i)) = 0$ can only be satisfied
if an element of
the set
\begin{align}
S_6 = \{ u\,,
v\,,
w\,,
1{-}u\,,
1{-}v\,,
1{-}w\,,
\frac{1}{u}\,,
\frac{1}{v}\,,
\frac{1}{w} \}
\label{eq:hexagonsingularities}
\end{align}
vanishes.  We conclude that the three-loop $n=6$ wheel graph has
first-type Landau singularities
on the locus
\begin{align}
\mathcal{S}_6 \equiv \bigcup_{s \in S_6} \{s = 0\}\,.
\label{eq:hexagonlocus}
\end{align}

It is straightforward, if somewhat tedious, to analyze all
\emph{subleading} Landau singularities corresponding to
relaxations, as defined above.
We refer the reader to~\cite{Dennen:2015bet,Dennen:2016mdk,Prlina:2017tvx}
where this type of analysis has been carried out in detail in
several examples.
We find no additional first-type singularities
beyond those that appear at leading order.
Let us emphasize that this unusual feature does not occur for any of the
examples in~\cite{Dennen:2015bet,Dennen:2016mdk,Prlina:2017tvx},
which typically have many additional subleading singularities.

To summarize, we conclude that any six-particle amplitude in any
four-dimensional massless planar field theory, at any finite loop
order, can have first-type Landau singularities only on the locus
$\mathcal{S}_6$
given by Eqs.~(\ref{eq:hexagonsingularities})
and~(\ref{eq:hexagonlocus}), or a proper
subset thereof.

\section{Second-Type Singularities}

The first-type Landau singularities
that we have classified, which by definition are those
encapsulated in the Landau equations~(\ref{eq:onshell}), (\ref{eq:kirk}),
do not exhaust
all possible singularities of amplitudes in general quantum
field theories.  There also exist
``second-type'' singularities (see
for example~\cite{Fairlie:1962-1,ELOP})
which are
sometimes called ``non-Landauian''~\cite{Cutkosky:1960sp}.
These arise in Feynman loop integrals as pinch singularities
at infinite loop momentum and must be analyzed by a modified
version of Eqs.~(\ref{eq:onshell}), (\ref{eq:kirk}).

In the next section we turn our attention to the
special case of SYM theory,
which possesses a remarkable \emph{dual conformal
symmetry}~\cite{Drummond:2006rz,Alday:2007hr,Drummond:2008vq}
implying that there is no invariant notion of ``infinity''
in momentum space.  As pointed out in~\cite{Dennen:2015bet},
we therefore expect that second-type singularities should be
absent in any dual conformal invariant theory.
Because ziggurat graphs are manifestly
dual conformal invariant when $D=4$,
this would imply that the first-type Landau singularities
of the ziggurat graphs should capture the entire
``dual conformally invariant part'' of the singularity
structure of all massless planar theories in four spacetime dimensions.
By this we
mean, somewhat more precisely, the singularity loci that do not
involve the infinity twistor.

\section{Planar SYM Theory}
\label{sec:sym}

In Sec.~\ref{sec:reduction} we acknowledged that in certain special
theories, the actual set of singularities of amplitudes may be strictly
smaller than that of the ziggurat graphs due to cancellations.
SYM theory has been shown to possess
such rich mathematical structure that it would seem the most
promising candidate to exhibit such cancellations.
Contrary to this expectation, we now argue that:
\vspace{0.05cm}
\begin{center}
\boxed{
\begin{minipage}{0.95\linewidth}
Perturbative amplitudes in SYM theory exhibit
first-type Landau singularities on
\emph{all} such loci that are possible in any massless
planar field theory.
\end{minipage}
}
\vspace{0.05cm}
\end{center}

Moreover, our results suggest that this all-order statement is true
separately in each helicity sector.
Specifically:
for any fixed $n$
and any $0 \le k \le n - 4$, there is a finite value of
$L_{n,k}$ such that the singularity locus
of the $L$-loop $n$-particle N${}^k$MHV amplitude
is identical to that of the
$n$-particle ziggurat graph for all $L \ge L_{n,k}$.
In order to verify this claim, it suffices to
construct an $n$-particle on-shell diagram with N${}^k$MHV
support that has the same Landau singularities as the $n$-particle
ziggurat graph; or (conjecturally) equivalently, to write down
an appropriate valid configuration of lines inside the
amplituhedron~\cite{Arkani-Hamed:2013jha}
$\mathcal{A}_{n,k,L}$ for some sufficiently high $L$.

To see that this is plausible, note that
in general the appearance of a given singularity at some
fixed $k$ and $L$ can be shown to imply the existence of the same
singularity
at lower $k$ but higher $L$ by
performing the opposite of a parallel reduction---doubling
one or more edges
of the relevant Landau graph to make bubbles (see for example
Fig.~2 of~\cite{Prlina:2017tvx}).
For example, while one-loop MHV amplitudes do not have singularities
of three-mass box type, it is known by explicit
computation~\cite{CaronHuot:2011ky}
that two-loop MHV amplitudes do.
Similarly, while two-loop MHV amplitudes do not have
singularities of four-mass box type, we expect that
three-loop MHV and two-loop NMHV amplitudes do. (To be clear,
our analysis is silent on the question of whether the
symbol alphabets
of these amplitudes contain square roots; see the discussion in
Sec.~7 of~\cite{Prlina:2017azl}.)

It is indeed simple to check that the $n$-particle ziggurat graph
can be converted into a valid on-shell diagram with MHV
support by
doubling \emph{each} internal edge to form a bubble.
Moreover, in this manner it is relatively simple to write
an explicit mutually positive configuration of positive
lines inside the MHV amplituhedron.
However, we note that while this construction is sufficient
to demonstrate the claim, it is certainly overkill; we expect MHV
support to be reached at much lower loop level than this argument
would require, as can be checked on a case by case basis for relatively
small $n$.

\section{Symbol Alphabets}

Let us comment on the connection of our work to symbol
alphabets.
In general, the presence of some letter $a$ in the symbol
of an amplitude indicates that there exists some sheet on which the
analytically continued amplitude has a branch cut from
$a=0$ to $a=\infty$.
The symbols of
all known six-particle amplitudes
in SYM theory can be expressed in terms of a nine-letter
alphabet~\cite{Goncharov:2010jf} which may be chosen
as~\cite{Dixon:2011pw}
\begin{align}
A_6 = \{ u, v, w, 1{-}u, 1{-}v, 1{-}w, y_u, y_v, y_w \}\,,
\label{eq:hexagonalphabet}
\end{align}
where $z = \{ y_u, 1/y_u\}$ are the two roots of
\begin{align}
\label{eq:quadratic}
u(1{-}v)(1{-}w)(z^2{+}1) = \left[ u^2{-}2 u v w{+}(1{-}v{-}w)^2 \right] z
\end{align}
and $y_v$ and $y_w$ are defined by cycling $u \to v \to w \to u$.
It is evident from Eq.~(\ref{eq:quadratic}) that $y_u$ can attain
the value $0$ or $\infty$ only if $u = 0$ or $v = 1$ or $w = 1$.
We therefore see that the singularity locus encoded in the hexagon
alphabet $A_6$ is precisely equivalent to $\mathcal{S}_6$
given by Eqs.~(\ref{eq:hexagonsingularities})
and~(\ref{eq:hexagonlocus}).
Indeed, the hypothesis that six-particle amplitudes in SYM
theory do not exhibit
singularities on any other loci
at any higher loop order (which we now consider to be
proven), and the apparently much stronger ansatz that
the nine quantities shown in Eq.~(\ref{eq:hexagonalphabet})
provide a symbol alphabet for all such amplitudes,
lies at the heart of a bootstrap program that has made possible
impressive explicit computations to high loop order
(see for example~\cite{Dixon:2011pw,Dixon:2011nj,Dixon:2013eka,Dixon:2014xca,Dixon:2014iba,Dixon:2015iva,Caron-Huot:2016owq}).
An analogous ansatz for $n=7$
has similarly allowed for the computation of symbols of seven-particle
amplitudes~\cite{Drummond:2014ffa,Dixon:2016nkn}.

Unfortunately, as the $y_u, y_v, y_w$ letters
demonstrate, the connection between Landau singularity loci and
symbol alphabets is somewhat indirect.
It is not possible to derive $A_6$ from $\mathcal{S}_6$ alone
as knowledge of the latter only tells us about the locus where symbol
letters
vanish~\cite{Maldacena:2015iua} or have branch points (see
Sec.~7 of~\cite{Prlina:2017azl}).
In order
to determine what the symbol letters actually are away from
these loci it seems necessary to invoke some other kind of structure;
for example, cluster algebras may
have a role to play here~\cite{Golden:2013xva,Drummond:2017ssj}.

\section{Conclusion}

We leave a number of open questions for future work.
What is the minimum loop order $L_n$ to which the $n$-particle
ziggurat graph can be reduced?
Can one characterize its Landau singularities
for arbitrary $n$, generalizing the result for $n=6$ in
Sec.~\ref{sec:landau}?
Does there exist a similar framework for classifying second-type
singularities, even if only in certain theories?
The graph moves reviewed in Sec.~\ref{sec:moves} preserve
the (sets of solutions to the) Landau equations even for non-planar
graphs;
are there results on non-planar $Y$-$\Delta$
reducibility (see for example~\cite{Wagner:2015,Gitler:2017})
that may be useful for non-planar (but still massless) theories?

In Sec.~\ref{sec:landau} we saw that the wheel is a rather remarkable graph.
The ziggurat graphs, and those to which they can be reduced,
might warrant further study for their own sake.
Intriguingly they generalize those studied
in~\cite{Bourjaily:2017bsb,Bourjaily:2018ycu} and are
particular cases
of the graphs that have attracted recent
interest, for example in~\cite{Chicherin:2017cns,Basso:2017jwq},
in the context of ``fishnet'' theories.
We have only looked at their singularity loci; it would be interesting
to explore the structure of their cuts, perhaps in connection
with the coaction studied
in~\cite{Abreu:2014cla,Abreu:2017ptx,Abreu:2017enx,Abreu:2017mtm,Abreu:2018sat}.

In the special case of SYM theory the technology might
exist to address more detailed questions.
For general $n$ and $k$, what is the minimum loop order
$L_{n,k}$ at which the Landau singularities of the $n$-particle
N${}^k$MHV amplitude saturate?
Is there a direct connection between Landau singularities,
ziggurat graphs, and cluster algebras?
For amplitudes of generalized polylogarithm type,
now that we know (in principle) the relevant singularity loci,
what are the actual symbol letters for general $n$, and
can the symbol alphabet
depend on $k$ (even though the singularity
loci do not)?
How do Landau singularities manifest themselves in general
amplitudes that are of more complicated (non-polylogarithmic) functional type?

\begin{acknowledgments}

We are grateful to C.~Colbourn for correspondence,
to N.~Arkani-Hamed for stimulating discussions,
to J.~Bourjaily for helpful comments on the draft,
and to T.~Dennen, J.~Stankowicz and A.~Volovich for collaboration
on closely related work.
We are especially indebted to
I.~Gitler for sending us the relevant portion of his Ph.D.\ thesis.
This work was supported in part by the US Department of Energy under
contract {DE}-{SC}0010010 Task A and
the Simons Fellowship Program in Theoretical Physics (MS).
\end{acknowledgments}

\appendix

\end{document}